\documentstyle[12pt,fleqn]{book}
\begin{document}
\hbox{}

\thispagestyle{empty}

${}$

\noindent
{\Large \bf G. Giachetta\footnote{Department of
Mathematics and Informatics, University of Camerino,
Italy} \newline
\newline L. Mangiarotti\footnote{Department of
Mathematics and Informatics, University of Camerino,
Italy}
\newline\newline G. Sardanashvily\footnote{Department of
Theoretical Physics, Moscow State University, Russia}}
\vskip3cm

\begin{center}

{\huge Geometric and Algebraic Topological 
\bigskip
\bigskip

 Methods in Quantum Mechanics}

\vskip7cm

{\large \bf World Scientific 
\bigskip
\bigskip

2005}

\end{center}

\newpage

\thispagestyle{empty}

${}$
\vskip2cm

\noindent
{\huge \bf Preface}  
\bigskip
\bigskip

\bigskip
\bigskip

Contemporary quantum mechanics meets an explosion of different
types of quantization. Some of these quantization techniques
(geometric quantization, deformation quantization, BRST
quantization, noncommutative geometry, quantum groups, etc.) call
into play advanced geometry and algebraic topology. These
techniques possess the following main peculiarities.

$\bullet$ Quantum theory deals with infinite-dimensional manifolds
and fibre bundles as a rule.

$\bullet$ Geometry in quantum theory speaks mainly the algebraic
language of rings, modules, sheaves and categories.

$\bullet$ Geometric and algebraic topological methods can lead to
non-equivalent quantizations of a classical system corresponding
to different values of topological invariants.

Geometry and topology are by no means the primary scope of our
book, but they provide the most effective contemporary schemes of
quantization. At the same time, we present in a compact way all
the necessary up to date mathematical tools to be used in studying
quantum problems.

Our book addresses to a wide audience of theoreticians and
mathematicians, and aims to be a guide to advanced geometric and
algebraic topological methods in quantum theory. Leading the
reader to these frontiers, we hope to show that geometry and
topology underlie many ideas in modern quantum physics. The
interested reader is referred to extensive Bibliography spanning
mostly the last decade. Many references we quote are duplicated in
{\it E-print arXiv} (http://xxx.lanl.gov).

With respect to mathematical prerequisites, the reader is expected to
be familiar with the basics of differential geometry of fibre bundles.
For the sake of convenience, a few relevant mathematical topics are
compiled in Appendixes.

\newpage

${}$
\vskip1.7cm

\noindent
{\huge \bf Contents}  
\bigskip
\bigskip\bigskip

\contentsline {chapter}{\numberline {Preface}}{v}
\contentsline {chapter}{\numberline {Introduction}}{1}

\contentsline {chapter}{\numberline {1}Commutative
geometry}{17}
\contentsline {section}{\numberline {1.1} Commutative
algebra}{17}
\contentsline {section}{\numberline {1.2} Differential
operators on modules and rings}{23}
\contentsline {section}{\numberline {1.3} Connections on
modules and rings}{27}
\contentsline {section}{\numberline {1.4} Homology and
cohomology of complexes}{31}
\contentsline {section}{\numberline {1.5} Homology and
cohomology of groups and algebras}{39}
\contentsline {section}{\numberline {1.6} Differential
calculus over a commutative ring}{56}
\contentsline {section}{\numberline {1.7} Sheaf cohomology}{59}
\contentsline {section}{\numberline {1.8} Local-ringed
spaces}{70}
\contentsline {section}{\numberline {1.9} Algebraic
varieties}{85}

\contentsline {chapter}{\numberline {2}Classical Hamiltonian
systems}{91}
\contentsline {section}{\numberline {2.1} Geometry and
cohomology of Poisson manifolds}{91}
\contentsline {section}{\numberline {2.2} Geometry and
cohomology of symplectic foliations}{110}
\contentsline {section}{\numberline {2.3} Hamiltonian
systems}{115}
\contentsline {section}{\numberline {2.4} Hamiltonian
time-dependent mechanics}{136}
\contentsline {section}{\numberline {2.5} Constrained
Hamiltonian systems}{157}
\contentsline {section}{\numberline {2.6} Geometry and
cohomology of K\"ahler manifolds}{172}
\contentsline {section}{\numberline {2.7} Appendix. Poisson
manifolds and groupoids}{189}

\contentsline {chapter}{\numberline {3}Algebraic
quantization}{195}
\contentsline {section}{\numberline {3.1} GNS construction I.
$C^*$-algebras of quantum systems}{195}
\contentsline {section}{\numberline {3.2} GNS construction
II. Locally compact groups}{209}
\contentsline {section}{\numberline {3.3} Coherent states}{217}
\contentsline {section}{\numberline {3.4} GNS construction
III. Groupoids}{224}
\contentsline {section}{\numberline {3.5} Example. Algebras
of infinite quibit systems}{229}
\contentsline {section}{\numberline {3.6} GNS construction
IV. Unbounded operators}{234}
\contentsline {section}{\numberline {3.7} Example. Infinite
canonical commutation relations}{238}
\contentsline {section}{\numberline {3.8} Automorphisms of
quantum systems}{249}

\contentsline {chapter}{\numberline {4}Geometry of algebraic
quantization}{257}
\contentsline {section}{\numberline {4.1} Banach and Hilbert
manifolds}{257}
\contentsline {section}{\numberline {4.2} Dequantization}{271}
\contentsline {section}{\numberline {4.3} Berezin's
quantization}{274}
\contentsline {section}{\numberline {4.4} Hilbert and
$C^*$-algebra bundles}{278}
\contentsline {section}{\numberline {4.5} Connections on
Hilbert and $C^*$-algebra bundles}{282}
\contentsline {section}{\numberline {4.6} Example. Instantwise
quantization}{286}
\contentsline {section}{\numberline {4.7} Example. Berry
connection}{290}

\contentsline {chapter}{\numberline {5}Geometric
quantization}{295}
\contentsline {section}{\numberline {5.1} Leafwise geometric
quantization}{295}
\contentsline {section}{\numberline {5.2} Example. Quantum
completely integrable systems}{306}
\contentsline {section}{\numberline {5.3} Quantization of
time-dependent mechanics}{312}
\contentsline {section}{\numberline {5.4} Example. Non-adabatic
holonomy operators}{324}
\contentsline {section}{\numberline {5.5} Geometric
quantization of constrained systems}{332}
\contentsline {section}{\numberline {5.6} Example. Quantum
relativistic mechanics}{335}
\contentsline {section}{\numberline {5.7} Geometric
quantization of holomorphic manifolds}{342}

\contentsline {chapter}{\numberline {6}Supergeometry}{347}
\contentsline {section}{\numberline {6.1} Graded tensor
calculus}{347}
\contentsline {section}{\numberline {6.2} Graded differential
calculus and connections}{352}
\contentsline {section}{\numberline {6.3} Geometry of graded
manifolds}{358}
\contentsline {section}{\numberline {6.4} Lagrangian
formalism on graded manifolds}{366}
\contentsline {section}{\numberline {6.5} Lagrangian
supermechanics}{382}
\contentsline {section}{\numberline {6.6} Graded Poisson
manifolds}{385}
\contentsline {section}{\numberline {6.7} Hamiltonian
supermechanics}{388}
\contentsline {section}{\numberline {6.8} BRST complex of
constrained systems}{392}
\contentsline {section}{\numberline {6.9} Appensix.
Supermanifolds}{401}
\contentsline {section}{\numberline {6.10} Appendix. Graded
principal bundles}{423}
\contentsline {section}{\numberline {6.11} Appendix. The
Ne'eman--Quillen superconnection}{426}

\contentsline {chapter}{\numberline {7}Deformation
quantization}{433}
\contentsline {section}{\numberline {7.1} Gerstenhaber's
deformation of algebras}{433}
\contentsline {section}{\numberline {7.2} Star-product}{444}
\contentsline {section}{\numberline {7.3} Fedosov's
deformation quantization}{450}
\contentsline {section}{\numberline {7.4} Kontsevich's
deformation quantization}{459}
\contentsline {section}{\numberline {7.5} Deformation
quantization and operads}{472}
\contentsline {section}{\numberline {7.6} Appendix. Monoidal
categories and operads}{475}

\contentsline {chapter}{\numberline {8}Non-commutative
geometry}{483}
\contentsline {section}{\numberline {8.1} Modules over
$C^*$-algebras}{484}
\contentsline {section}{\numberline {8.2} Non-commutative
differential calculus}{486}
\contentsline {section}{\numberline {8.3} Differential
operators in non-commutative geometry}{492}
\contentsline {section}{\numberline {8.4} Connections in
non-commutative geometry}{498}
\contentsline {section}{\numberline {8.5} Connes'
non-commutative geometry}{503}
\contentsline {section}{\numberline {8.6} Landsman's
quantization via groupoids}{507}
\contentsline {section}{\numberline {8.7} Appendix.
$K$-Theory of Banach algebras}{509}
\contentsline {section}{\numberline {8.8} Appendix. The
Morita equivalence of $C^*$-algebras}{512}
\contentsline {section}{\numberline {8.9} Appendix. Cyclic
cohomology}{514}
\contentsline {section}{\numberline {8.10} Appendix.
$KK$-Theory}{518}

\contentsline {chapter}{\numberline {9}Geometry of quantum
groups}{523}
\contentsline {section}{\numberline {9.1} Quantum groups}{523}
\contentsline {section}{\numberline {9.2} Differential
calculus over Hopf algebras}{530}
\contentsline {section}{\numberline {9.3} Quantum principal
bundles}{535}

\contentsline {chapter}{\numberline {10}Appendixes}{541}
\contentsline {section}{\numberline {10.1} Categories}{541}
\contentsline {section}{\numberline {10.2} Hopf algebras}{546}
\contentsline {section}{\numberline {10.3} Groupoids and Lie
algebroids}{553}
\contentsline {section}{\numberline {10.4} Algebraic Morita
equivalence}{565}
\contentsline {section}{\numberline {10.5} Measures on
non-compact spaces}{569}
\contentsline {section}{\numberline {10.6} Fibre bundles I.
Geometry and connections}{586}
\contentsline {section}{\numberline {10.7} Fibre bundles II.
Higher and infinite order jets}{611}
\contentsline {section}{\numberline {10.8} Fibre bundles III.
Lagrangian formalism}{618}
\contentsline {section}{\numberline {10.9} Fibre bundles IV.
Hamiltonian formalism}{626}
\contentsline {section}{\numberline {10.10} Fibre bundles V.
Characteristic classes}{633}
\contentsline {section}{\numberline {10.11} $K$-Theory of
vector bundles}{648}
\contentsline {section}{\numberline {10.12} Elliptic
complexes and the index theorem}{650}

\contentsline {chapter}{\numberline {Bibliography}}{661}

\contentsline {chapter}{\numberline {Index}}{683}

\end{document}